# Detection of Gravitational Lenses and Measurement of Time Delays from Radiation Fluctuations

Ermanno F. Borra,

Centre d'Optique, Photonique et Laser,

Département de Physique, Université Laval, Québec, Qc, Canada G1K 7P4

(email I: borra@phy.ulaval.ca)

# ABSTRACT

I suggest that measurements of intensity fluctuations caused by classical wave interactions can be used to find unresolved gravitational lenses and determine time delays of essentially arbitrary length among the images formed by a gravitational lens. No interferometry is needed, the time delays can be measured by analyzing the intensity signal alone. The technique works with lensed sources that have constant luminosities and is capable of measuring very long time delays. I suggest interferometric techniques, capable of measuring time delays of arbitrary length, that can be used to refine the measurements.

## 1. INTRODUCTION

Following the discovery of a gravitational lens by Walsh, Carswell, & Weymann (1979) interest in the phenomenon has exploded and there is now a vast literature dealing with it (Blandford, & Naryan 1992). Time delays are predicted among the signals from the images of a gravitational lens that can be exploited for a variety of studies (Blandford, & Naryan 1992). Determining them by observing delays in luminosity variations, as has been considered in the past, is time consuming, restricts one to resolved lenses that have variable sources and, with the exception of rare and yet undiscovered lensed pulsars or gamma-ray bursters, to measurements of long delays. This article points out that intensity fluctuations caused by classical wave interactions carry a lensing signature that can be used to identify unresolved lenses and measure time delays. No interferometry is needed, the time delays can be measured by analyzing the intensity signal alone. The theory and applications of radiation fluctuations are reviewed by Mandel (1962a). I briefly discuss techniques, complementary to the present one, for the accurate interferometric measurement of time delays of arbitrary length.

## 2. RADIATION FLUCTUATIONS AND TIME DELAYS

Consider a pulse V(t) and its frequency spectrum $G(\omega)$ given by the Fourier transform of V(t); the shift theorem gives the frequency spectrum of a pulse shifted in time by $\tau$ as $e^{-i\omega\tau}G(\omega)$. Superposition of two identical pulses shifted by $\pm \tau/2$ then yields the frequency spectrum

$$H(\omega) = 1/(2\pi)^{1/2} \int_{-\infty}^{+\infty} [V(t+\tau/2) + V(t-\tau/2)] e^{-i\omega t} dt \qquad (1)$$

from which one obtains, with the shift theorem,

$$H(\omega) = 2\cos(\omega\tau/2)G(\omega) \quad , \tag{2}$$

a result known as the converse modulation theorem.

The frequency spectrum is now modulated by a $\cos(\omega\tau/2)$ term that allows to distinguish a signal due to V(t) alone from a signal obtained from V(t+$\tau$/2) + V(t-$\tau$/2). These textbook results (Bracewell 1986) give the bases of the method by which we can determine the time delay $\tau$ in a gravitational lens since an unlensed object would have spectrum G($\omega$) and an lensed one H($\omega$). A similar discussion was used by Alford & Gold (1957) to devise an ingenious method to measure the speed of light. For brevity, we assume double images of equal intensities and model a gravitational lens as a two-beam interferometer: the theory can readily be extended to include more complex situations. However, we now face two apparent difficulties: most objects are not pulsed and modulation of the spectral density will be impossible to determine with a spectrograph, for equation 2 predicts that the minima are spaced by $\Delta\omega = 2\pi\Delta n/\tau$ and are too close together to be resolved for all but tiny values of $\tau$. Both difficulties can be overcome. First, we can consider a photon as a wave packet V(t) that is then split in two by an interferometer that then adds a time delay to one of the pulses: the analysis leading to Eq. 2 thus applies. Givens (1961) reconsidered the Alford & Gold experiment as a wave propagation problem showing that the frequency spectrum I($\omega$) of the current I(t) = $<|V(t)|^2>$ is given by

$$I(\omega) = (2\pi)^{1/2} \cos(\omega\tau/2) \int_{-\infty}^{+\infty} a(\omega'+\omega) a^*(\omega') G(\omega'+\omega) G^*(\omega') d\omega' \qquad (3)$$

where a(ω) describes the bandpass of the detector and the asterisk indicates complex conjugation. I(t) is the instantaneous current that would be measured at the output of a square law detector; Mandel (1962a) discusses the measurement of radiation fluctuations with a finite time resolution. A more rigorous relation for $I^2$, including the effect of shot noise, has been obtained by Mandel (1962b) but equation 3 is less cumbersome to manipulate and visualize and, since it illustrates the salient features of the effect just as well, we shall use it for discussion purposes. Equation 3 shows that the frequency spectrum of I(t) is modulated by cos(ωτ/2) and, as Givens (1961) and Mandel (1962b) point out, does not depend on pulsing so that the experiment can be carried out with unmodulated light, as long as the two beams are coherent. The effect is thus directly applicable to the lensing of objects that are not pulsed. Pulsation was required in the Alford & Gold experiment to impose a correlation to two otherwise uncorrelated beams. Second, as in Alford & Gold (1957), we can measure the current I(t) and detect the spectral minima of its fluctuations that, as we shall see, appear at much lower frequencies than those of H(ω).

For astronomical applications, the bandpass is small compared to its central frequency so that we can assume

$$a(\omega')G(\omega') = g \exp(-(\omega'-\omega_0)^2/\sigma^2), \qquad (4)$$

allowing us to evaluate the integral in Equation 3 in closed form as

$$I(\omega) = K \cos(\omega\tau/2) \exp(-1/2\,\omega^2/\sigma^2), \qquad (5)$$

where I have grouped all the constants in K after making use of the fact that the error function, that appears after performing the integral in equation 3 with equation 4, tends to $\pm 1$ for $\omega'$ tending to $\pm \infty$. Equation 5 shows that $I(\omega)$ is non-zero at arbitrarily low frequencies and is modulated with periods $= 2\tau/(2n+1)$. It illustrates the key effect for it shows that spectral modulation occurs at frequencies well outside the spectral bandpass of the light measured. It allows the measurements of time delays in gravitational lenses for which $\tau$ can range from less than $10^{-8}$ seconds for microlenses to over a year for objects lensed by massive objects. It must also be noted, as discussed at length by Givens (1961), that *the above discussion is valid for arbitrarily long values of $\tau$, independently of the coherence time given by the bandpass*: The experiment can be done with broadband radiation.

Equation 5 can be intuitively understood by considering that the instantaneous signal fluctuations are due to wave beats among all the frequencies in the light beam, much as wave beats modulate the carrier frequency in a radio detector. A lensed source differs from an unlensed one because its frequency spectrum $H(\omega)$ is modulated by $\cos(\omega\tau/2)$ (e.g., some frequencies are missing because the two beams have path differences having an odd number of half wavelengths); hence the beats of a spectrally modulated and an unmodulated source should be expected to have different power spectra. Since beats induce lower frequencies equal to the difference among the beating frequencies, the closer the beating frequencies, the lower will the frequency of the fluctuation be. This allows one to carry out very high resolution spectroscopy (Mandel 1962a) so that the extremely fine spectral features predicted by Eq. 2 for long $\tau$ become detectable as current fluctuations of much lower frequency.

The discussion so far has neglected the effect of noise. The kind and magnitude of noise to consider will depend on the instruments and techniques used, which will also depend on the region of the electromagnetic spectrum, and is beyond the scope of this paper. In principle instrumental or background noise can be nearly eliminated but photon shot noise is a source of noise that cannot be eliminated and gives a fundamental limit that we can readily estimate. Mandel (1962b) discusses the effect of shot noise. From his work, we can write the ratio R between the wave-interaction and shot noise terms of the spectral densities of the fluctuations as

$$R \leq \delta = \alpha \bar{I} / \Delta \nu. \qquad (6)$$

where $\delta$ is the degeneracy parameter (Mandel 1962b) and where I have used a convenient measure of $\delta$ obtained from an expression given by Purcell (1956) and adopted $1/\Delta\nu$ as a convenient measure of the coherence time, where $\Delta\nu = \Delta\omega/2\pi$ is the bandpass of observation. The parameter $\alpha$ represents the quantum efficiency of the system so that $\alpha I$ is the average photon counting rate. Equation 6 shows that shot-noise dominates, making the effect hard to detect, whenever $\delta \ll 1$, namely when the counting rates are below one count per coherence time interval, while it becomes easier to detect if $\delta > 1$. Shot noise is further considered in the discussion.

Other source of noise come from interstellar scintillation and background gravitational waves: they are considered in the discussion.

## 3. INTERFEROMETRIC MEASUREMENTS OF TIME DELAYS: A COMPLEMENTARY TECHNIQUE

The time delays obtained from the technique can be improved with more sensitive interferometric techniques. Mandel (1962b) has reexamined the classical superposition experiment such as occurs in a Young or Michelson interferometer. Mandel(1962b) shows that there is spectral modulation of the normalized spectral density $\Phi_{33}(v)$ at the output given by

$$\Phi_{33}(v) = \Phi_{11}(v)\{1 + [2(\bar{I}_1\bar{I}_2)^{1/2}/(\bar{I}_1 + \bar{I}_2)]\gamma_{12}(0)\cos(2\pi vT)\}, \qquad (7)$$

where, keeping the notation of Mandel, $\Phi_{11}(v)$ is the normalized input spectral density, T the time delay, $\gamma_{12}(0)$ the normalized crosscorrelation function at T=0 and $v$ the frequency of light. It must be noted that *spectral modulation occurs for steady beams and for arbitrarily long path differences cT*, as pointed out by Mandel (1962b). *There is no contradiction with the usual textbook statements that no interference occurs for OPDs greater than the coherence length.* It is indeed is the case for intensity modulation (Mandel, 1962b) treated in textbooks, but it is not for spectral modulation. Givens (1961) explains spectral modulation from several points of view, but an easy way to understand it is by considering that some frequencies are missing because they have path differences having an odd number of half wavelengths: Spectral modulation must occur.

The spectral modulation due to a time delay is beyond the resolution of astronomical spectrographs for OPDs greater than a few meters in the visible and a few kilometers in the radio region. However there is a technique that can measure path differences arbitrarily large (Brochu, & Delisle 1972). Cielo, Brochu, & Delisle (1975) demonstrated it with a path difference of 300 meters and a white light source. Basically, the technique consists in observing the output of the first interferometer with a second interferometer identical to the first one; actually Cielo, Brochu, & Delisle (1975) used the same interferometer in double pass. Note that the second interferometer sees the output $\bar{I}_3$ of the first one so that the gravitational lens needs not be resolved. The method however assumes a prior knowledge of the path differences, greatly reducing its usefulness; but it can be adapted. Cielo and Delisle (1977) have proposed a communication system that modulates the signal by varying the path length of one of the interferometers. They show that if the path lengths of the two interferometers are not equal, the Intensity $\bar{I}_6$ output by the second interferometer is given by

$$\bar{I}_6 = (\bar{I}_4 + \bar{I}_5)[1 + \alpha^2/2 \int \cos[2\pi v(T_2 - T_1)]\Phi_{11}(v)dv], \qquad (8)$$

where $\bar{I}_4$ and $\bar{I}_5$ are the mean intensities of the two beams of the second interferometer, $T_2$ -$T_1$ is the difference between the time delays in the two interferometers and, assuming $\gamma_{12}(0) = \gamma_{45}(0)$,

$$\alpha = 2(\bar{I}_4 \bar{I}_5)^{1/2} \gamma_{45}(0)/(\bar{I}_4 + \bar{I}_5) \qquad . \qquad (9)$$

Equation 8 shows that the signal $\bar{I}_6/(\bar{I}_4 + \bar{I}_5)$ is modulated by the Fourier cosine transform of $\Phi_{11}(v)$. Let us assume that the source has a featureless flat continuum and that the observations are made with a square bandpass, an adequate representation of an interference filter or a radio astronomical bandpass, so that the spectral distribution seen by the interferometer is of the form $\Phi_{11}(v)$ = constant for $v0 - \Delta v/2 < v < v0 + \Delta v/2$ and $\Phi_{11}(v) = 0$ outside these boundaries. Assuming $\gamma_{12}(0) = \gamma_{45}(0) = 1$ and with $\bar{I}_4 = \bar{I}_5$ we obtain $\alpha = 1$ and upon integration

$$\bar{I}_6/(\bar{I}_4 + \bar{I}_5) = 1 + \cos[2\pi(T_2 - T_1)v_0]\sin[2\pi(T_2 - T_1)\Delta v/2]/2\pi(T_2 - T_1)\Delta v \quad . \qquad (10)$$

Setting $\gamma_{12}(0) = \gamma_{45}(0) = 1$, we assume perfect spatial coherence, which is far from being valid for the known gravitational lenses. This strong assumption will be examined in the discussion. Equation 10 predicts a strongly modulated signal having an envelope that decreases in amplitude with ΔT. It has modulated sidelobes having an envelope that decreases as $1/\Delta T \Delta \nu$. A non-rectangular bandpass or a continuum with strong features or a substantial slope will give a different decrease. Working at longer wavelengths or using a narrower bandpass stretches the signal, rendering it more easily detectable.

Once an approximate time delay is found by direct measurements of the intensity fluctuations, it could be improved and its time variations monitored with the interferometric techniques suggested by this section. One could determine the interferometric signature by physically changing the path difference in an interferometer. A more versatile approach borrows from radio astronomy and very-long-baseline interferometry where the signals are recorded with separate telescopes to be correlated later (Cohen 1969). It is therefore possible to record independently the individual signals from separate antennas, perform the correlations leading to equation 8 with trial values of $T_2$ and look for the lensing signature

Note that, in principle, one could identify a lens with the interferometric techniques without an a priori knowledge of the time delay by simply changing the path difference in an interferometer until $\bar{I}_6 /(\bar{I}_4 + \bar{I}_5)$ deviates sufficiently from unity. Unfortunately, this direct method is inefficient for all but very small values of $T_1$. Furthermore one is limited by the practical difficulties of building an interferometer capable of very large OPDs.

The very-long-baseline interferometric technique could, in principle, measure arbitrarily large values of $T_1$ without the a priori knowledge of the time delays. In practice, however, one will be limited by the time needed to perform the numerous trial correlations: This will be a major problem for delay times, or uncertainties in the delay times greater than a few hours.

The direct approach has been suggested by Spillar (1993), as pointed out by an anonymous referee of the original version of this article. However, the approach used in the present article, that starts from the spectral density, has the advantage of showing that the time delays can be found irrespective of the time correlation length. It therefore predicts that one can find time delays of arbitrary length, hence lenses of arbitrary mass. Spillar's article is only concerned with the short time delays given by sub-solar mass lenses. Actually, the short time delays considered by Spillar (1993) can easily be found by observing the spectral modulation predicted by Eq. 7.

## 4. DISCUSSION

Section 2 assumes a perfectly spatially coherent source. The more rigorous treatment of Mandel (1962b) includes spatial coherence. He shows that the signal is proportional to $\gamma_{12}^2(0)$, the coefficient of spatial mutual coherence. Schneider & Schmid-Burgk (1985) have studied the mutual coherence of images of QSOs gravitationally lensed by a point source, finding that $\gamma_{12}(0) \ll 1$ for masses of practical interest, rendering the effect totally beyond detection. This can also be readily seen by considering that a typical lensed QSO (Blandford & Naryan 1992), can be thought as being observed by an interferometer having "arms" separated by a=10 Kpc. Applying the vanCittert-Zernicke theorem, we find that at wavelength $\lambda$, spatial coherence requires that the angle $\theta$ subtended by the source be such that $\theta < \lambda/a$, therefore for $\lambda$=6cm one needs $\theta < 10^{-20}$. At a distance of 1 Gpc, this corresponds to dimensions $<10^5$ cm, several orders of magnitude smaller than realistic source sizes. Because the "separation" of the arms scales as the square root of the mass of the lens, it is difficult to find astronomically relevant combinations of mass, distances and luminosities.

Prima facie, coherence considerations would seem to render this technique hopelessly inefficient. However, Mandzhos (1991a,b) has examined the mutual coherence properties of images of quasars observed through gravitational lenses that have complicated structures, finding that the mutual coherence can increase by orders of magnitudes with respect to lensing by a single point source. In particular (Mandzhos 1991b) finds that the mutual coherence function for a quasar microlensed by a binary star can approach unity. Binary stars constitute a large fraction of the stellar population of a galaxy so that the probability of lensing by a binary star is not negligible (Mandzhos 1991b). It would thus appear that time delays could be detected by observing radio-loud QSOs and looking for microlens time scales. Other compact sources could also be observed (e.g. pulsars in our galaxy or extragalactic masers). Mandzhos (1996) has extended his original analysis to consider issues specific to this work, indicating that his original theory must be extended to be useful for a quantitative interpretation of the effects here discussed.

Photon shot noise gives a fundamental limit for the detection of the effect so that we can estimate at what fluxes the effect is detectable. The degeneracy parameter $\delta$ (Eq. 6) is hopelessly small in the optical region of the spectrum for all but the brightest astronomical sources and largest telescopes since $h\nu$ is large and the coherence times small. The degeneracy parameter $\delta$ increases with decreasing frequency so that one will do better in the radio region. To estimate the fluxes at which shot noise becomes important we can use equation 6. Assuming $\nu = 1$ GHz and $\alpha=0.5$ we obtain $\delta=230$ for a 1 Jansky source observed with a 100-m diameter radiotelescope. This indicates that gravitational lenses could be detected at the mJ level with large radiotelescopes if other sources of noise could be eliminated. Extraneous noise could be minimized by observing with two nearby telescopes and admitting only the fluctuations that are correlated: $\delta$ does not depend on the bandpass since the increase in counts $\sim \Delta\nu$ is compensated by the $1/\Delta\nu$ factor in Eq. 6. One can obtain a larger value of $\delta$ at lower frequencies since $\delta \sim 1/\nu$ for a given flux. The measurement of angular sizes and structure of radio sources with intensity interferometry (Hanbury Brown, Jennison, & Das Gupta 1952; Jennison, & Das Gupta 1956) demonstrates that wave-interaction effects are indeed detectable in astronomical radio sources.

Since the signal is modulated by the low beat frequency $\nu_1$, the effect is valid for $\Delta\nu T \gg 1$ and the coherence condition $\delta\nu T$, with $\delta\nu$ the bandpass of the low frequency observations, replaces the usual $\Delta\nu T$ coherence condition (Mandel 1962b). This signifies that, like in the intensity interferometer experiment (Hanbury Brown 1968), one can observe inside a large bandpass $\Delta\nu$ and that the requirements of the optical quality of the mirrors used are vastly more relaxed than for conventional interferometry so that poor quality but large reflectors can be used. One could therefore build large inexpensive low surface quality specialized telescopes to gather large fluxes, allowing to overcome the coherence and degeneracy problems.

Equation 5 shows that there is little power for $\omega > 2\sigma$. In fact, there will be none at high frequencies for the assumption of a Gaussian bandpass (Eq. 4) breaks down for $(\omega - \omega_0)^2 \gg \sigma^2$ since it predicts, contrarily to facts, some contribution at arbitrarily large or low $\omega$. If we had used a rectangular bandpass $\Pi[\omega - \omega_0]$ instead of a Gaussian, the autocorrelation in Equation 3 would have given the triangle function $\Lambda[\omega]$ with no power at all at large $\omega$. This puts a fundamental restriction on the time delays that can be measured at a given frequency. Consider that the bandpass must be such that, for an unambiguous identification of the lensing signature, there must be at least n zeros from the $\cos(\omega\tau/2)$ term in equation 5. It can readily be seen that we must have

$$\tau > (2n+1)\pi/\sigma. \qquad (11)$$

Since $\sigma < \omega_0$, we must also have

$$\omega_0 > (2n+1)\pi/\tau. \qquad (12)$$

For example, observing at 10 GHz with a 0.1 GHz bandpass would restrict us to $\tau > 2 \cdot 10^{-7}$ seconds if we require at least 10 zeros for an unambiguous detection of the signature. This corresponds to a microlens of about $10^{-2}$ solar masses. Weaker lenses, that induce shorter time delays would have to be measured at higher frequencies, where unfortunately $\delta$ is smaller. Note that there is no restriction to the region of the electromagnetic spectrum we can work at; so that, subject to the flux limitations imposed by the degeneracy parameter, one could work in the visible with $\sigma \sim 10^{13}$ and measure $\tau > 10^{-12}$ seconds. These orders of magnitude estimates obviously neglect limitations imposed by the instrumentation.

We can consider two regimes of detection depending on the value of $\tau$. For delay times longer than the time resolution of the instrumentation one can determine the power spectrum $I(\omega)$ by Fourier analysis of the signal $I(t)$. The difficulty is that the amplitude of the spectral modulation is small and there will be noise from macroscopic flux variations of the source as well as various sources of internal (e.g., variations in the gain of the detectors) and external noise. The effect however carries a well-defined signature that could be detected by using standard techniques to extract a signal from a noisy background (e.g., correlation techniques). For values of $\tau$ comparable or smaller than the resolution time of the instrument one could determine directly $I(\omega)$ by sending the $I(t)$ signal from the telescope output to a sharply tuned electrical filter using a simple setup similar to those described by Alford & Gold (1957), Givens (1961) and Mandel (1962b). Note that the Alford & Gold (1957) experiment was done with a photomultiplier in the visible region of the spectrum, illustrating that measurements of small time delays ($10^{-7}$ seconds in that experiment) can be carried out and that one can work at all frequencies.

We have glossed over a few additional complications. For example, gravitational lenses have beams of unequal strength and may generate multiple images; but the theory can readily be generalized to handle such cases. Also, the signal will be washed out if the lensing geometry changes, inducing variations of the time delays. On the positive side, modeling these complications will give information on the source and the lensing event. Compact extragalactic sources have an extended component that will introduce background noise. An additional complication may come from interstellar scintillation that introduces delays between adjacent rays, masquerading as gravitational lens delays. Scintillation effects can be minimized by working at high frequencies; unfortunately our analysis shows that low frequencies are better suited. Observing at high galactic latitudes will also lessen scintillation noise. Scintillation effects can however be separated from gravitation effects because the former are frequency-dependent, while the latter are not. Note also that, as some time one person's garbage is another person's treasure, the technique may be used to study the interstellar medium. These complications shall eventually have to be studied quantitatively, with an emphasis on the signal to noise ratio of the detections.

A discussion of astronomical applications of the technique is beyond the scope of this letter, but we can consider that detection of a lensing signature from the modulation of the spectrum of the fluctuations (e.g., Equation 5) will give an unequivocal confirmation of lensing that can be used to find unresolved lenses. Measuring time delays is of obvious general interest since it allows measurements of the lensing geometry and the geometry of spacetime. A particularly interesting possibility concerns the detection of gravitational waves (Allen 1989), in particular the background of primeval waves that some cosmological theories predict. Gravitational lenses can be thought as interferometers in space: A passing gravitational wave changes the geometry of space-time, inducing an additional time delay. We could thus conceivably use the technique to detect gravitational waves by measuring the derivative of $\tau$ in a gravitational lens. The challenge, however, will consist in separating the intrinsic delay of the lens from the additional delay introduced by the passing wave. Conceivably, one could use $d\tau/dt$ to discriminate between changes in the geometry of the lens and the signature of a gravitational wave. Interstellar scintillation effects could be removed by observing at two wavelengths.

5. **CONCLUSION**

Measurements of classical radiation fluctuations can be used to find unresolved gravitational lenses and determine time delays among the images formed by a gravitational lens. The effect is however difficult to detect because it is highly dependent on the spatial coherence of the source. It is hopelessly small for astronomical objects lensed by a simple lens (e.g. point source) but becomes detectable for complex lenses (e.g. double stars). Coherence considerations favor work at radio wavelengths rather than the optical region. Photon shot noise also degrades the effect and favors long wavelengths.

It is unfortunate that this technique is greatly hampered by spatial coherence effects but, fortunately, like the Hanbury-Brown -Twiss intensity interferometer, it is far less sensitive to temporal coherence effects than classical interferometry. This means that one can use telescopes having far less accurate surfaces than suggested by the wavelength the receptor works at. Coherence considerations suggest that a search for gravitational lenses be carried out at long radio wavelengths with very large radio telescopes. Because of the relaxed surface quality requirements, one could construct large inexpensive Arecibo-like radio telescopes dedicated to such a search. The only requirement on the surface quality of the antenna is that its point spread function be small enough that background noise from the sky and nearby sources be small. Considering that the observations would be confined to very bright compact sources, one could get away with very poor surfaces indeed. Considering the scientific pay-back one could get by measuring delay time (e.g. detection of gravitational waves), it should be well-worth the expense and effort of building such instruments. The highly successful search for MACHOs illustrates the kind of pay-back one can get from a large dedicated search for rare events. Once a candidate lens has been found and the delay time roughly determined, detailed quantitative study is best left to the interferometry technique described in section 3.

The main purpose of this paper is to point out that the energy distributions of lensed sources carry information that can be used to detect and measure the lensing event; therefore several complications have been glossed over (e.g. scintillation noise, lens geometry) and shall eventually have to be addressed quantitatively.

REFERENCES

Alford, W.P., & Gold, A. 1958, Am. J. Phys., 26, 481

Allen, B, 1989 Phys. Rev. Letters 63, 2017

Blandford, R.D., & Naryan, R. 1992, ARA&A, 30, 311

Bracewell, R.N. 1986, The Fourier Transform and its Applications (New York: McGrawiHill)

Brochu, M. & Delisle, C. 1972, Can. J. Phys., 50,1993

Cielo, P., Brochu, M. & Delisle, C. 1975, Can J. Physics, 53, 1047

Cielo, P. & Delisle, C. Can J. Phys. 1977, 55, 1221

Cohen, M.H. 1969, ARA&A, 7, 619

Givens M.P. 1961, J. Opt. Soc. Am., 51,1030

Hanbury Brown, R., Jennison, R.C., & Das Gupta, M.K. 1952, Nature, 170, 1061

Hanbury Brown, R. 1968, ARA&A, 6,13

Jennison, R.C., & Das Gupta, M.K. 1956, Phil. Mag. Ser.8, 1, 65

Mandel, L. 1959, Proc. Phys. Soc. (London) 74, 233

Mandel, L. 1962a, Progress in Optics, 2, 183

Mandel, L. 1962b, J. Opt. Soc. Am., 52, 1335

Mandzhos, A.V. 1991a, Sov. Astron. 35, 11

Mandzhos, A.V. 1991b, Sov. Astron. 35, 116

Mandzhos, A.V. 1996, private communication

Purcell, E.M. 1956, Nature 178, 1449

Schneider, P., & Schmid-Burgk, J. 1985, A&A 148, 369

Spillar, E.J. 1993 ApJ 403, 20

St-Arnaud, J.M. & Delisle, C. 1970, Can J. Phys., 48, 2112

Walsh, D., Carswell, R.F., & Weymann, R.J. 1979, Nature, 279, 381

1 9ACKNOWLEDGMENTS

This research was supported by a grant from the Natural Sciences and Engineering Research Council of Canada. I wish to thank Dr. A. Mandzhos for useful comments on an earlier version of this manuscript. I thank an anonymous referee for bringing Spillar's article to my attention and for comments and suggestions that helped to improve the original manuscript